\documentclass[pre,twocolumn,showpacs,amsmath,amssymb]{revtex4}

\usepackage{graphicx}
\usepackage{dcolumn}
\usepackage{bm}

\begin{document}
\title{A mechanism for air induced fluidization in vibrated granular beds}
\author{L. I. Reyes}
\author{I. S\'anchez}
\author{G. Guti\'errez}
\affiliation{Departamento de F\'{\i}sica, Universidad Sim\'on Bol\'{\i}var, 
Apartado 89000, Caracas 1080-A, Venezuela}

\date{\today}
\pacs{45.70.-n, 47.55.Kf, 45.70.Mg, 64.70.-p} 
\keywords{granular matter, solid-fluid transitions, density segregation}

\begin{abstract}

 We present a new mechanism for the fluidization of a vibrated granular 
bed in the presence of interstitial air. We show that the air flow induced across the bed, 
as the gap in the bottom of the cell evolves, 
can fluidize the bed in a similar way as in gas-fluidized {\it static} beds. 
We use the model of Kroll to quantify the relevant variables suggested by the mechanism 
proposed. The relevance of the above fluidization for segregation phenomena is discussed.
\end{abstract}

\maketitle

A forced collection of grains can exhibit very rich and striking behavior 
that sometimes resembles the properties of fluids and solids \cite{Duran,Jaeger}. 
Fluidlike behavior can be induced, among other means, by shaking 
\cite{Jaeger,Vibration} or by injecting gas through the grains \cite{fluid,Menon,Valverde}.
Due to its particular transport properties gas-fluidized granular beds
have found many aplications in industry \cite{fluid}. Vibration commonly induces
undesirable segregation in many industrial processes which involve the handling of 
large collections of grains \cite{Duran,Jaeger}. From a fundamental point of view, 
forced granular beds offers a plethora of interesting collective behavior. 
Thus a basic understanding of phase changes as a response of these dissipative systems 
to a change in forcing conditions, grains
properties and of any relevant parameter is of practical and fundamental interest.

The influence of interstitial air in phenomena related to vibrated granular media have been
studied in different contexts \cite{Kroll,Gutman,Air1,Air2,Naylor,Yan,comp,Dust}.
Under certain conditions, interstitial air has proved to be a crucial ingredient in
segregation phenomena \cite{Air2,Naylor,Yan,comp,Dust},
and in this line some efforts to include air in molecular dynamics simulations have 
appeared recently \cite{comp}. However, a coherent description that would allow us to 
understand the role played by the air in vibrated granular systems is still lacking. 

In this article we present a mechanism by which a vibrated bed of grains
can become fluidized in the presence of interstitial air. 
The key ingredient
is not just the presence of air but the way it flows through the vibrated bed. 
We discuss the consequences of this kind of fluidization in terms of a previously 
proposed quantitative model \cite{Gutierrez} for reverse buoyancy \cite{Shinbrot}. 

We assume that the vibrated grain pack acts like a porous piston so the acceleration
of the container induces a pressure difference between its ends as the pack moves relative
to the bottom of the container, which in turn produces an air flow across the bed.
We use Kroll's model \cite{Kroll} to describe the evolution of the gap under the grain pack,
within one period of vibration.
This model takes into account a force that 
arises as a consequence of the pressure difference across the bed that develops as 
the gap in the bottom evolves. We propose that the flow of air induced by this pressure 
difference can fluidize the medium in much the same way as it occurs in gas-fluidized 
{\it static} beds, where gas is injected against gravity. The model of Kroll will allow 
us to present explicit relations between the flow of air through the bed, the pressure 
difference between its ends and the gap formed at the bottom of the bed.
From these relations we develop a quantitative description of the proposed mechanism
of fluidization.\\

\noindent
{\it Kroll's Model}\\

\begin{figure}[t]
\begin{center}
\includegraphics[width=5cm]{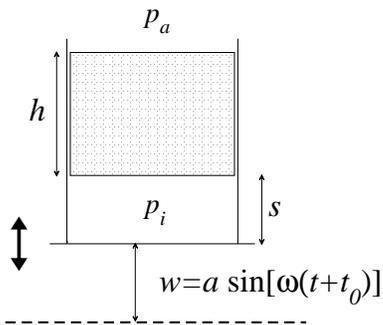}
  \caption{\label{esquema} The Kroll's picture: the granular bed is treated as a porous piston.
The container is subjected to sinusoidal vibrations of frequency $\omega$ and amplitude $a$.
At the top of the bed we have atmospheric pressure $p_a$ and $p_i$ is the time dependent 
pressure in the gap.}
\end{center}
\end{figure}

In the Kroll's model \cite{Kroll} the granular bed its treated as a porous piston. 
When the effective gravity starts to point upward the piston takes off.
According to Kroll's model, in the noninertial reference frame placed on the container, 
only two forces are taken into account during the flight: the effective weight of the granular
bed and the force due to the pressure difference across the piston.
The following evolution equation for the gap at the bottom of the bed is obtained 
(see figure \ref{esquema}):
\begin{equation}\label{eq1}
\frac{d^2s}{dt^2}+(p_a-p_i)\frac{A}{m}=-\frac{d^2w}{dt^2}-g,
\end{equation}
where $s$ is the size of the gap, $p_a$ is the atmospheric pressure, 
$p_i$ is the time dependent pressure in the gap, 
$A$ is the transversal area of the container, $m$ is the mass of the piston, 
$g$ is the acceleration of gravity and $w$ is the position as a function of time of 
the container. 

We can use an equation of state to relate
the gap, and hence the volume of the air in the gap, with the pressure $p_i$.
The air is assumed to behave as an incompressible fluid \cite{Kroll,Gutman}.
This results in the additional equation
\begin{equation}\label{eq2}
\frac{ds}{dt}=\frac{1}{\rho_a A}\frac{dG}{dt},
\end{equation}
where $G$ is the mass of air in the gap and $\rho_a$ is the density of air. 
It is assumed that the flux of air throught the bed is given by Darcy's law 
\cite{Kroll,Sahimi} which gives us the following equation 
\begin{equation}\label{eq3}
\frac{dG}{dt}=\rho_a A\left[ \frac{k}{\mu} \frac{(p_a-p_i)}{h}\right],
\end{equation}
where $k$ is the permeability of the piston, $h$ it's height and
$\mu$ is the viscosity of air. 
Equations (\ref{eq1})-(\ref{eq3}) can be integrated for the unknowns $s(t)$, $p_i(t)$ 
and $G(t)$.  We assume that the container is subjected to sinusoidal oscillations of 
amplitud $a$ and angular frequency $\omega$, so we have that $w=a\sin[\omega(t+t_0)]$,
where $\sin(\omega t_0)=1/\Gamma$ and $\Gamma=a\omega^2/g$. At $t=0$ the medium takes off. 

If we introduce the adimensional quantities $s'=s/a$ and $t'=\omega t$, 
we find the following evolution equation for the adimensional gap $s'$:
\begin{equation}\label{evol}
\ddot{s'}+\frac{1}{\tau_K}\dot{s'}=\frac{1}{\Gamma}[\Gamma\sin(t'+t_0')-1],
\end{equation}
where $\dot{s'}=ds'/dt'$, $t_0'=\omega t_0$ and the adimensional number $\tau_K$ is given by
\begin{equation}\label{Omega}
\tau_K=\frac{\omega k \rho_m}{\mu}.
\end{equation}

We see that in this model the effect of air in the piston movement is a usual frictional 
factor, with a friction coefficient given by $1/\tau_K$. We expect then that if $\tau_K$ 
is small then the gap will be small. 
The adimensional number $\tau_K$ depends on the grains diameter through the permeability $k$ 
of the medium \cite{Sahimi}. 
An interesting prediction that arises is that the gap's 
dynamics depends only on the adimensional numbers $\Gamma$ and $\tau_K$. 
Integrating equation (\ref{evol}) with initial 
conditions $\dot{s'}(0)=s'(0)=0$ we obtain the evolution of the adimensional gap $s'(t')$:
\begin{eqnarray}\label{sigma}
 s' &  = & \frac{\tau_K}{\Gamma} [ C -t' 
-\tau_K( 1+\lambda\Gamma \sin\delta )e^{-t'/\tau_K}\nonumber \\
&   &  -\lambda\Gamma\cos(t'-\delta)] ,
\end{eqnarray}
where $C=\tau_K+\lambda\Gamma\left(\tau_K\sin\delta+\cos\delta\right)$,
$\lambda=\left(1+\tau_K^2\right)^{-1/2}$, and
$\delta=\arctan \tau_K - \arcsin 1/\Gamma$. Equation (\ref{sigma}) is valid while $s'>0$.\\

The gap is small \cite{coment} when a large pressure difference between the ends of the bed is 
induced. From equations (\ref{eq2}) and (\ref{eq3}) we see that in this model the pressure 
difference across the bed is proportional to the first derivative of $s$ and is given by 
\begin{equation}\label{presion}
\frac{p_a-p_i}{gh\rho_m}=\frac{\Gamma}{\tau_K}\dot{s'}.
\end{equation}

From Darcy's law, the flow of air through the grains is proportional to this pressure 
difference.\\

\noindent
{\it Gas-Fluidized beds and cyclic fluidization in vibrated granular beds}\\

In the technique of fluidizing beds by injecting gas against gravity 
a fluidized state is achieved when the force due to the pressure difference across the bed is high enough 
to cancel the weight of the bed \cite{fluid,Menon,Valverde}. An important control parameter 
in this technique is the velocity $U$ at which gas is injected upward at the 
bottom of the {\it static} bed. As the gas flows up through the granular medium it induces a pressure
difference that results in an effective force on the bed that point upward. This force rises 
as $U$ is increased from zero and when it equals the weight of the bed fluidlike behavior 
starts to be observed \cite{fluid,Menon}. This fluidization condition can be written as
\begin{equation}\label{fluidC}
\Delta P A\ge mg,
\end{equation}
where $\Delta P$ is the pressure difference between the ends of the bed.

In the Kroll's model the pressure difference
and the effective gravitational field are time dependent. This can be seen
if we rewrite the evolution equation (\ref{eq1}) for the gap as
\begin{equation}\label{eq4}
m\frac{d^2s}{dt^2}=mg_{ef}(t)-[p_a-p_i(t)] A,
\end{equation}
where $g_{ef}(t)=g\{\Gamma \sin[\omega (t+t_0)]-1\}$ is the effective gravitational field.
Before taking off, when $g_{ef}$ points downward and 
the bed is in contact with the bottom of the container, there is no significant air flow,
the bed is being pushed downward against the bottom of the container
and there is no relative movement between the grains; 
under these conditions we consider the bed as being
in a solid state. When $g_{ef}$ starts to point upward the bed takes off, 
the size of the gap and the air flow through the grains starts to increase. 
Following condition (\ref{fluidC}), for this case we can write 
the fluidization condition as follows ($g_{ef}\ne 0$):
\begin{equation}\label{fluidUS}
\frac{[p_a-p_i(t)] A}{mg_{ef}(t)}\ge 1.
\end{equation}

\begin{figure}[t]
\begin{center}
\includegraphics[width=8cm]{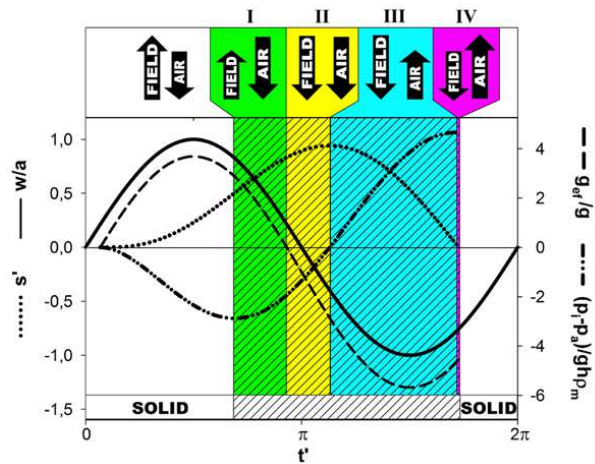}
\caption{\label{untitled} (Color online) Within a period of vibration, we show in this figure 
the evolution in time of the following adimensional variables: the gap $s'$ (from eq. (\ref{sigma})),
the pressure difference across the bed $(p_i-p_a)/gh\rho_m$ (from eq. (\ref{presion})), 
the effective gravitational field acting on the bed $g_{ef}/g$ 
and the position of the container $w/a$ for the experimental 
conditions of reference \cite{Gutierrez}.
The begining of the fluidized state occurs at the inflection point of the curve $s'$ vs $t'$.
It is shown in which fraction of the cycle the bed is considered to be in a solid state.
Within the fraction of the cycle that runs from $t'_f$ until the bed hits the 
bottom of the cell at the end of its flight, we identify four zones depending
on the direction and intensity of the effective gravitational field and of the air flow. 
The direction of these quantities is shown at the top of the figure.}
\end{center}
\end{figure}

Combining equations (\ref{eq4}) and (\ref{fluidUS}), 
the fluidization condition (\ref{fluidUS}) is initially satisfied at the inflection 
point of the curve $s'$ vs $t'$ shown in figure \ref{untitled}. 
This result is independent of the validity of equations (\ref{eq2}) and (\ref{eq3})
and, consequently, of the assumptions made in deriving them.
If we assume that equations (\ref{eq2}) and (\ref{eq3}) holds, 
the beginning of the fluidized state occurs at the adimensional time $t'_f$ (measured
from the time when $g_{ef}$ starts to point upwards) which is given by
\begin{equation}\label{fluidCond}
\cos(t'_f-\delta)e^{t'_f/\tau_K}=\frac{1+\lambda\Gamma\sin\delta}{\lambda\Gamma\tau_K}.
\end{equation}

Within the Kroll's assumptions, at $t'_f$ the
pressure difference across the bed is maximum.
From equations (\ref{evol}) or (\ref{presion}) and the discussion above,
it is expected that the mechanism proposed will be effective if $\tau_K$ is small.
Equation (\ref{fluidCond})) gives, to our knowledge, a new result. With this equation
we can estimate when the new mechanism of fluidization is initiated. This can be experimentally
verified by measuring what happens to the granular medium at the inflection point of the
curve that represents the evolution of the gap. This, for example, predicts for the
case of reverse buoyancy, a change in the motion of an intruder at time $t'_f$.

In figure \ref{untitled} we show the evolution in time of the gap (from eq. (\ref{sigma})), 
the pressure difference across the bed (from eq. (\ref{presion})), 
the effective gravitational 
field acting on the bed and the position of the container for the experimental conditions 
of reference \cite{Gutierrez}. Within the fraction of the cycle that runs from $t'_f$ 
until the bed hits the bottom of the cell at the end of its flight,
we can identifie four zones depending on the direction and intensity of the effective
gravity $g_{ef}$ and of the air flow. The fluidization condition (\ref{fluidUS}) 
is satisfied in zones {\bf I} and {\bf IV} (see figure). 
If the flight time of the bed is small compared with the period of vibration, 
then we have a granular bed that alternates cyclically between a solid and a fluidized state. \\

\begin{figure}[ht]
\begin{center}
\includegraphics[width=6.5cm]{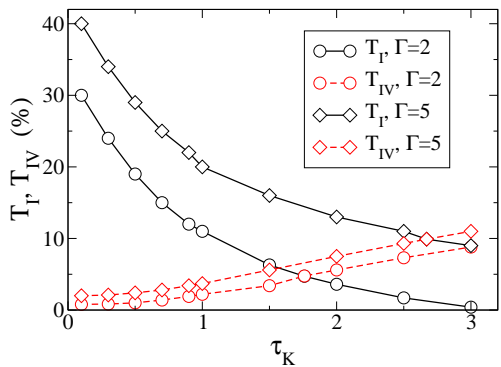}
  \caption{\label{Tfases}(Color online) Time that the bed will spend in zone {\bf I} ($T_I$) 
and in zone {\bf IV} ($T_{IV}$) of figure \ref{untitled} as a function of $\tau_K$ calculated 
from the model of Kroll, for $\Gamma=2$ and $\Gamma=5$. 
$T_I$ and $T_{IV}$ are reported as its porcentual fraction with respect to 
$T_I+T_{II}+T_{III}+T_{IV}$, that is the time in which an intruder immersed in the bed can move with respect to
the small grains. For $\tau_K>3$ the flight time of the bed is close to the period of vibration.}
\end{center}
\end{figure}

\noindent
{\it Cyclic Fluidization and Reverse Buoyancy}\\

We are now going to explore the consecuences of the above mechanism of fluidization in 
the context of the phenomenon of reverse buoyancy. Under certain conditions large 
heavy objects immersed in a vibrated granular medium rise and similar light ones sink 
to the bottom. This was called reverse buoyancy \cite{Shinbrot}. 
In reference \cite{Gutierrez} it was proposed that this phenomenon could be
explained by assuming that the granular bed behaves as a fluid in only a fraction of 
the cycle of vibration. The necessary feature is that during most of the time in which 
the bed is in a fluidized state the effective gravity points upward. 

In the fluidization condition explained in the previous section, 
the air flow points downward just after the bed takes off, while  
the effective gravity points upward \cite{coment2}. 
If a buoyancy force arises in the above fluidized state, large heavy objects immersed in it may
{\it sink} upward while similar light ones may {\it float} downward.
This case is portrayed in zone {\bf I} shown in figure \ref{untitled}.
In zone {\bf IV} the fluidization condition (\ref{fluidUS}) is also
satisfied, but now $g_{ef}$ is pointing downwards and we expect to observe regular buoyancy 
in this zone. Further work is needed to anticipate how the bed would behave in zones 
{\bf II} and {\bf III}, but it is expected that relative movement between the grains is 
posible in these zones. In particular, we can ask ourselves: 
is there a buoyancy force in zones {\bf II} and {\bf III}?. 
In this line, valuable information can be obtained by resolving the motion
of an intruder immersed in the bed within a period of vibration and comparing it
with the evolution of the measured pressure difference across the bed. 
Other mechanisms can play a role in these zones \cite{Dust}.

The time that the bed would spend in each of the zones shown in figure \ref{untitled} 
will depend on experimental conditions. Within the model of Kroll, it would depend on 
$\Gamma$ and $\tau_K$. If reverse buoyancy is observed, then it could be a sign 
that zone {\bf I} prevails. In figure \ref{Tfases} we show the time that the bed would
spend in zones {\bf I} ($T_I$) and {\bf IV} ($T_{IV}$) as predicted by the model of Kroll. It can be seen
that if $\tau_K$ is small then we have an scenario that favors reverse buoyancy.
The model predicts that there is a particular value $\tau_{K,c}(\Gamma)$ for which
$T_I=T_{IV}$. For $\tau_K=\tau_{K,c}$ the average motion of an intruder immersed in the bed 
would depend on how the bed behaves in zones {\bf II} and {\bf III}.

In reference \cite{Yan} reverse buoyancy was supressed by evacuating the air of the system.
If we quench the flow of air through the bed for the experimental conditions of reference 
\cite{Gutierrez} by keeping the top of the container open and introducing a filter permeable 
to air in it's bottom, as was made in \cite{Naylor}, we do not observe reverse buoyancy: 
light and heavy intruders rise. In both procedures the proposed fluidization mechanism 
cannot take place.

In this article we have presented a mechanism by which a vibrated granular bed can be fluidized 
in the presence of interstitial air in much the same way as it occurs in gas-fluidized 
{\it static} beds. This mechanism results from an interplay between the flow of air through 
the bed and the effective gravitational field acting on the bed.
In order to quantify the relevant variables of our model, 
we presented explicit results for the model of Kroll.
Since the mechanism proposed is an extension of an {\it static} one, and since the bed cannot switch between a 
fluidized state and a solid state instantly,
we expect that this results are aplicable when the period of vibration is large.
Under this condition, we can predict when a vibrated granular system initially becomes
gas fluidized during each cycle of oscillation. The proposed model constitutes a promising 
quantitative framework that gives some relevant information about the process of cyclic fluidization
that occurs in a vibrated granular bed \cite{Gutierrez}. It would be important to monitor 
experimentally the evolution of the granular medium, within one period of oscillation of the system.


\begin{thebibliography}{99}
\bibitem{coment} If the granular packing undergoes ballistic flight it can be shown that 
for $\Gamma > \Gamma_c\approx 1.81$ the maximum gap would be greater than the amplitude of vibration $a$. 
So for $\Gamma > \Gamma_c$ if $s'_{max}$ is small then we can say that the gap is small.
If $\tau_K$ is small and $\Gamma$ is large we find that the maximum adimensional gap is given by
$s'_{max} \simeq \left(\frac{2\Gamma -\pi}{\Gamma}\right) \tau_K$.
\bibitem{coment2} It can be shown that within the model
of Kroll the fluidization condition (\ref{fluidUS}) always occurs
when the effective gravity $g_{ef}$ is still pointing upwards.
\end{thebibliography}
\end{document}